\documentclass[preprint]{JHEP}
\usepackage{epsfig}
\relax

\newcommand\fverb{\setbox\pippobox=\hbox\bgroup\verb}
\newcommand\fverbdo{\egroup\medskip\noindent%
                        \fbox{\unhbox\pippobox}\ }
\newcommand\fverbit{\egroup\item[\fbox{\unhbox\pippobox}]}
\newbox\pippobox

\usepackage{amssymb}
\usepackage{amsmath}

\newcommand{\be}{\begin{equation}}
\newcommand{\ee}{\end{equation}}
\newcommand{\ba}{\begin{align}}
\newcommand{\ea}{\end{align}}
\newcommand{\bea}{\begin{eqnarray}}
\newcommand{\eea}{\end{eqnarray}}
\newcommand{\bw}{\begin{widetext}}
\newcommand{\ew}{\end{widetext}}

\newcommand{\e}{{\rm e}}

\newcommand{\nn}{\nonumber}
\newcommand{\zt}{\dot{z}}

\newcommand{\y}{{\mathbf{y}}}

\title{Classical ultra-relativistic scattering in ADD}

\author{Dmitry V. Gal'tsov$^*$, Georgios Kofinas$^\dagger$, Pavel Spirin$^*$ and Theodore N.
Tomaras$^\dagger$\\

$^*$ Department of Theoretical Physics, Moscow State University,
119899, Moscow, RUSSIA\\
$^\dagger$ Department of Physics and Institute of Plasma Physics,
 University of Crete, 71003 Heraklion, GREECE\\

{\tt E-mail: galtsov@physics.msu.ru, gkofin@phys.uoa.gr,
salotop@list.ru, tomaras@physics.uoc.gr}}

\abstract{ The classical differential cross-section is calculated
for high-energy small-angle gravitational scattering in the
factorizable model with toroidal extra dimensions. The three main
features of the classical computation are: (a) It involves summation
over the infinite Kaluza-Klein towers but, contrary to the Born
amplitude, it is finite with no need of an ultraviolet cutoff. (b)
It is shown to correspond to the non-perturbative saddle-point
approximation of the eikonal amplitude, obtained by the summation of
an infinite number of ladder graphs of the quantum theory. (c) In
the absence of extra dimensions it reproduces all previously known
results.}

\begin{document}

\section{Introduction}

The search for large extra dimensions (LED), especially in view of
the forthcoming experiments at LHC, constitutes a very exciting
direction in high energy physics. After precursor ideas about the
Universe as a topological defect in higher-dimensional space-time
\cite{defect} and the proposal of TeV-scale internal dimensions
related to supersymmetry breaking in string theory \cite{ablt}, large
extra dimensions were discussed in several contexts. Among the
conceptually and technically simplest is the ADD picture \cite{ADD},
according to which the standard model particles live in a
four-dimensional space (the brane), embedded in a $D$-dimensional
bulk inhabited only by gravity, and with the extra $\delta=D-4$
dimensions compactified on a torus. The $D$-dimensional Planck mass
$M_*$ is supposed to lie in the TeV region and the LED have
submillimeter size. In this scenario there is an infinite tower of
massive Kaluza-Klein (KK) gravitons \cite{GRW} whose existence may
be detected at present and future colliders \cite{ABQ}, either as
missing energy in collisions due to emission of KK gravitons (which
are weakly interacting after being created), or via processes which
would otherwise be impossible or very much suppressed.

Quantum collision processes with exchange of virtual KK gravitons
are a very useful tool to test the model \cite{ABQ}. Unfortunately,
tree diagrams containing the propagator of the KK tower diverge in
the ultraviolet (UV) due to an infinite sum over the KK graviton
masses \cite{GRW}. The divergence appears already at {\em tree}
level and is due to the emission of infinite momenta into the
compactified dimensions. Several proposals were made in order to
cure this problem. One was to introduce a UV cutoff \cite{GRW,GiSt},
another to take into account brane oscillations and the associated
Nambu-Goldstone modes \cite{NG}, another yet to introduce brane
thickness \cite{GuSj}, or finally, to sum up ladder diagrams within
the eikonal approximation \cite{Em}. However, none of them solves
the problem completely. The UV cutoff at Planckian energies while it
certainly does exist, does not look very natural in this context. It
would mean that we are able to probe the Planckian regime using
(relatively) low energy processes. Appealing to features associated
with the {\em physical brane}, such as tension and thickness,
introduces additional elements into the ADD model, which seem to
spoil its overall consistency. Indeed, for a brane of non-zero
tension one can not use any more the flat Minkowski background.
Instead, one has to deal with solutions of the Einstein equations,
which are non-trivial even in the absence of matter on the brane. In
the case of codimension one, such a solution is well-known and is
the basis of the Randall-Sundrum II model \cite{RS}, with a
different graviton spectrum. Finally, the eikonal calculation
amounts to using the Fourier-transform of the Born amplitude and
thus, it also suffers from ambiguities associated with the
divergence of the latter. Indeed, the evaluation of the eikonal
phase gives a finite result if a certain order of integration is
used in calculating the Fourier integral, while it is divergent if
one merely changes the order of integration.

On the other hand, it has been argued that at ultrahigh energies,
particle scattering in four dimensions not only becomes dominated by
gravity, but in addition it involves only {\it classical}
gravitational dynamics \cite{TH,Mu,ACV,BaFi}. Indeed, quantum
gravity effects should not, by definition, be important in the
classical limit $\hbar \to 0$. This, in terms of the two relevant
lengths, i.e. the Planck length $l_{\rm{Pl}}=(\hbar G_4/ c^3)^{1/2}
=\hbar/M_{\rm{Pl}}c$ and the gravitational radius associated with
the energy of the collision $r_g=G_4 \sqrt{s} / c^4$, implies that
$r_g\gg l_{\rm{Pl}}$, which is equivalent to the condition
$\sqrt{s}\gg M_{\rm{Pl}} c^2$ of transplanckian energies. Thus, to
study the scattering of two point-like particles with Planck energy,
't Hooft used a shock wave approximation for the field of the moving
particle and obtained a result similar to the Veneziano amplitude.
Later on, it was shown that in four dimensional quantum gravity
\cite{KaOr}, as well as in string theory \cite{ACV}, the eikonal
approximation (free in both cases of the ambiguities mentioned
above) reproduced the result of 't Hooft.

The Planck length $l_*$ and the gravitational radius $r_g^*$ in the
$D$-dimensional ADD model are, correspondingly
 \be
 l_{*}=\left(\frac{\hbar G_D}{ c^3}\right)^{\frac1{\delta+2}}\sim
 \frac{\hbar}{M_{*}c},  \;\;\;\;  r^*_g=\left(\frac{G_D
 \sqrt{s}}{c^4}\right)^{\frac1{\delta+1}},
 \ee
where $M_{*}$ is the TeV-scale mass parameter. The above reasoning remains
essentially the same and shows that in the transplanckian regime
$\sqrt{s}\gg M_{*}c^2$ scattering is also classical, at least for
some range of momentum transfer. Moreover, the particularly exciting
proposals of black hole creation at LHC or in cosmic rays \cite{ADM}
\cite{MMT} were based on a purely classical picture.

Although the eikonal approximation of particle scattering in ADD has been discussed by a number of authors  \cite{Em}, \cite{GuSj}, no classical calculation of the cross-section was found in the literature. The purpose
of the present paper is to fill in this gap, and in addition to provide an independent check of the validity
of the eikonal approach, whose applicability in the context of ADD is not yet rigorously proven. Indeed,
it is hereby demonstrated explicitly that the classical theory reproduces the saddle-point result of the
eikonal approximation and is essentially non-perturbative in the quantum sense. Thus, classical theory
gives non-trivial results in ADD, in
contrast to four dimensional gravity where the classical elastic scattering cross-section coincides with the Born
approximation and {\em is} perturbative.
Finally, it should be pointed out that the reason the classical cross-section is finite, even though
it too involves an infinite summation over massive KK modes, is because in the classical
calculation the potentially divergent integrals contain oscillating factors, which effectively cutoff
the modes that cannot be excited by the source. This applies to all kinematical regimes and, in particular,
is analogous to the situation in the classical derivation of Newton's law \cite{KeSf} \cite{Sj}.

\section{The setup}
Consider the Fierz-Pauli lagrangian in $D$-dimensional Minkowski
space, with $\delta\equiv D-4$ of its spatial dimensions being a
torus $T^\delta$ with equal radii $R$ \be {\cal L}= -\frac{1}{4}
h^{MN} \Box h_{MN} +\frac{1}{4} h \Box h  -\frac12 h^{MN}\partial_M
\partial_N h  +\frac12 h^{MN}\partial_M
\partial_P h^P_N - \frac{\varkappa_D}{2}  h^{MN}T_{MN},
\label{FP} \ee where $M,N,..=0,1,2,...,D-1$. The Minkowski metric is
$\eta_{MN}={\rm diag} (1,-1,-1,...)$ and $\Box\equiv
\eta^{MN}\partial_M\partial_N$. The gravitational field $h_{MN}$ is
coupled to a conserved matter stress-tensor $T_{MN}$ ($\partial_N
T^{MN}=0$). All fields ($h\equiv \eta^{MN}h_{MN}\equiv h^M_M$) are
functions of \be x^M=(x^\mu,y^i),\quad  \mu =0,\dots ,3,\quad i=
1,\dots  \delta , \ee and are periodic under the translations
$y_j\to y_j +2\pi R, \quad j=1,\dots ,\delta $. Thus, for instance
\be
h_{MN}(x^P)=\sum_{n_1=-\infty}^{+\infty} \dots
\sum_{n_\delta=-\infty}^{+\infty} \frac{h^n_{MN}(x)}{\sqrt{V
}}\exp\left( i\frac{n_i y^i}{R}\right) ,
 \ee
where $  V =(2\pi R)^\delta $ is the volume of the torus. In what follows we abbreviate the sum over the KK modes as
$\sum_{n}$ and denote the momenta transversal to the brane as ${  p^i}_T=n^i/R.$ The tensor $h_{MN}(x^P)$
has been split into an infinite sum of four-dimensional KK modes $h^n_{MN}(x^\mu)$ with (mass)$^2$ equal to $p_T^2$.

In the harmonic gauge $\partial_N h^{MN}=\frac12 \partial^M h$
the Einstein equations for $h_{MN}$ read:
 \be
\Box h_{MN}=-\varkappa_D\left(T_{MN}-\frac{T}{\delta+2}\eta_{MN}
\right)\equiv -J_{MN}, \quad T=\eta^{MN}\,T_{MN}.
\label{EEQ}
\ee
According to the ADD scenario it is assumed that the matter stress-tensor is localized on the brane, located
at ${\bf y}=0$,
and carries only four-dimensional indices:
\be
T_{MN}(x^P)=\eta_M^\mu \eta_N^\nu T_{\mu\nu}(x) \delta^\delta(\y).
\label{ST}
\ee
It is worth noting, however, that this assumption is consistent with
Einstein's equations only at the linearized level, with matter dynamics
governed by the non-gravitational forces alone.
Given (\ref{ST}) it is consistent to set the graviphotons $h_{i\mu}$ and
the non-diagonal part of the scalar matrix $h_{ij}$ to zero.
Finally, the diagonal components of $h_{ij}$ are all equal and
generated by the trace of the energy-momentum tensor:
 \be
\Box h_{ij}=-\varkappa_D \frac{T}{\delta+2}\delta_{ij}.
 \ee
Their zero modes $n^i=0$ are the so called {\it radions}, which describe deformations of the
torus caused by the presence of matter on the brane. The massive
components $n^i\neq 0$ of $h^n_{MN}$ could be rearranged into the
massive four-dimensional graviton and massive scalars in the usual Higgs mechanism
language \cite{KT}, \cite{GRW}, but this is not necessary for the present discussion.

The $D$-dimensional Planck mass $M_*$ is defined by
 $\varkappa_D^2=16\pi/M_*^{\delta+2}\equiv 16\pi G_D$ and is related to the four-dimensional one $M_{\rm{Pl}}$ via
 $M_{\rm{Pl}}^2= M_*^{\delta+2}V$.

The retarded Green's function of the
D'Alembert equation satisfies
\begin{align}
\label{KKb0}
\Box  G_{D}(x,x',\y-\y')= - \delta^4(x-x')\delta^\delta(\y-\y').
\end{align}
Its Fourier transform reads:
\be \label{KK1a22_mi1}
 G_{D}(x-x',\y-\y') =\frac1{(2\pi)^4 V }
 \int d^4 p \, \e^{-ip\cdot (x-x')} \sum_n \frac{\e^{i{\bf p}_T\cdot (\y-\y')}}{p^2-p_T^2+i
\epsilon p^0}. \ee The solution of (\ref{EEQ}) with the source
localized on the brane \be J_{MN}(x,\y )=J_{MN}(x)\delta^\delta(\y),
\ee is \bea  h_{MN}(x,\y)&=&   \int
G_{D}(x-x',\y-\y')J_{MN}(x',\y')d^4 x' d\y'\nn\\
&=& \int \frac{d^4 p \, e^{-ip\cdot (x-x')}}{(2\pi)^4V} \,
\sum_n\frac{\e^{i{\bf p}_T\cdot \y} }{{p}^2-p_T^2+i \epsilon p^0 }\;
J_{MN}(x')\;d^4x'. \eea Its restriction to the  brane
 \be
h_{MN}(x)\equiv
h_{MN}(x,\y)\Big|_{\y=0}
 \ee
 can be rewritten using the amputated propagator
\begin{align}
\label{KKb2} h_{MN}(x)=\int G_{4}(x-x')J_{MN}(x') d^4 x',\quad
G_{4}(x-x')=G_{D}(x-x',\y-\y')\left.
\vphantom{\sqrt{f}}\right|_{\y=\y'=0}.
\end{align}
The  momentum space four-dimensional retarded propagator thus reads
\begin{align}
\label{KK1a22_mi1}
G_{4}(p)=\frac{1}{V }\sum_n \frac{1}{{p}^2-p_T^2+i \epsilon p^0}.
\end{align}
Equivalently, taking the four-dimensional Fourier transform, defined
by $\Psi(k)\!=\!\!\int \!\Psi (x) \e^{i k\cdot x} d^4 x$ and
$\Psi(x)\!=\!\int \Psi (k)\e^{-i k\cdot x} d^4 k/(2\pi)^4$
 of the above equations,
the retarded solution of Eq. (\ref{EEQ}) becomes
\be
h_{MN}(k)=\frac{\varkappa_D}{V}\sum_n\frac{T_{MN}(k)-\frac{1}
{\delta+2}\eta_{MN}T(k)}{{k}^2-p_T^2+i
\epsilon k^0}.
\label{ksol}
\ee

\section{The ultra-relativistic elastic scattering cross-section}
Consider next the small angle ultrarelativistic scattering of two
particles on the brane, with masses $m$ and $m'$ respectively,
interacting via $D$-dimensional gravity. Using for notational
simplicity the same parameter $\tau$ for both trajectories, a
convenient way to describe them is \be\label{zz}
z^\mu(\tau)=z^\mu_0+ \frac{p^\mu}{m}\tau+\delta z^\mu(\tau),\quad
z'^\mu(\tau)=z'^\mu_0+ \frac{p'^\mu}{m'}\tau+\delta z'^\mu(\tau),
\ee
 with $\delta z, \delta z'$ treated perturbatively.
The asymptotic values of their momenta are
\begin{align}
&   P^\mu\!=\!m\lim_{\tau\to-\infty}
\zt^\mu(\tau)=p^\mu\!+\!m\lim_{\tau\to-\infty}
\delta\zt^\mu(\tau),\,\,
  P'^\mu\!=\!m'\lim_{\tau\to-\infty}
\zt'^\mu(\tau)=p'^\mu+m'\lim_{\tau\to-\infty}
\delta\zt'^\mu(\tau),\nonumber\\& {\bar P}^\mu=m\lim_{\tau\to\infty}
\zt^\mu(\tau)=p^\mu+m\lim_{\tau\to\infty} \delta\zt^\mu(\tau),\quad
 {\bar P}'^\mu=m'\lim_{\tau\to\infty}
\zt'^\mu(\tau)=p'^\mu+m'\lim_{\tau\to\infty} \delta\zt'^\mu(\tau),
\end{align}
while momentum conservation implies $P^\mu+P'^\mu={\bar P}^\mu+{\bar
P}'^\mu$. The Mandelstam variables $s,\,t$ are
 \be
 s=(P+P')^2,\quad  t=q^2=({\bar P}-P)^2=m^2 \left(\lim_{\tau\to\infty}
 \delta\zt^\mu(\tau)-\lim_{\tau\to-\infty}
 \delta\zt^\mu(\tau)\right)^2, \label{tdef}
 \ee
and we consider the ultrarelativistic regime and the small-angle
approximation
  \be s  \gg m^2,\;m'^2,\; |t|,
\ee in which case $s=(p+p')^2$.

Since the momentum transfer $q^\mu$ depends only on the deviation
$\delta z^\mu$ of one of the particles, it will be convenient to
work in the rest frame of $m'$ before collision. In that frame
 \be
 p'^\mu=m'(1,0,0,0),\quad p^\mu=m\gamma(1,0,0,v),\quad
 \gamma=\frac1{\sqrt{1-v^2}},
 \ee
 and with no loss of generality one may set in addition
 \be
z'^\mu_0=0,\quad z^\mu_0=b^\mu=(0,b,0,0),
 \ee
 where $b$ is the impact parameter.

The particles equations  of motion following from the (\ref{FP}) are
$D$-dimensional geodesic equations in the metric
 \be
 g_{MN}=\eta_{MN}+\varkappa_D h_{MN},
 \ee but it is easy to show that the particle moving on the brane
 in zero order in $\varkappa_D$ will remain on the brane. So
the matter stress-tensor of the two particle system is
four-dimensional: \be T^{\mu\nu}=-\int \left[m \zt^\mu \zt^\nu
\delta^4(x-z(\tau))+ m' \zt'^\mu \zt'^\nu
\delta^4(x-z'(\tau))\right] d\tau. \label{stress} \ee

Leaving aside the classical mass renormalization needed to take into
account the gravitational self-action, we have to take as
 $h_{\mu\nu}$ in the equation for the  particle $m$ the retarded
field of the partner particle $m'$. In the small-angle approximation
it is assumed that the deviation from the unperturbed rectilinear
motion is small, and we get perturbatively
 \be
\Pi^{\mu\nu}\delta\ddot{z}_\nu=\varkappa_D
\Pi^{\mu\nu}\left(h_{\nu\lambda,\rho}-\frac12
h_{\lambda\rho,\nu}\right) \frac{p^\lambda p^\rho}{m^2},
\label{eqmot}
 \ee
 where $\Pi^{\mu\nu}$ is the projector onto the space transverse to the momentum $p^\mu$
 \be
 \Pi^{\mu\nu}=\eta^{\mu\nu}-\frac{p^\mu p^\nu}{m^2}.
 \ee

The gravitational field $h_{\mu\nu}$ of the particle $m'$ is given by (\ref{ksol}),
where we have to substitute the Fourier transform of the second term of the stress-tensor (\ref{stress})
\be
 h_{\mu\nu}(k)=\frac{2\pi \varkappa_D}{
 V}\sum_n\frac{\delta(k\cdot p')}{{k}^2-p_T^2+i \epsilon k^0}\left(p'_\mu
p'_\nu- \frac{m'^2\eta_{\mu\nu}}{ \delta+2} \right).
 \ee

Using the reparametrization invariance of the particle trajectories one may choose
$\tau$ in such a way that $p^\mu\delta \zt_\mu=0$ for both particles. In this gauge the solution of (\ref{eqmot})
becomes
 \be
 \delta z^\mu(\tau)=\frac{i\varkappa_D^2}{(2\pi)^3 V}\int\sum_n
 \frac{\delta(k\cdot p')\e^{ik\cdot b-ik\cdot p\tau/m}}{(k\cdot p)({k}^2-p_T^2+i\epsilon
 k^0)}Q^\mu d^4k.
 \ee
The vector $Q^\mu$ has the form
 \be
 Q^\mu=A k^\mu + B p^\mu +C p'^\mu,\label{Q}
 \ee
 where
 \be
A=\frac1{2k\cdot p}\left((p\cdot p')^2-\frac{m^2m'^2}{\delta+2}\right),\quad
B=\frac1{2m^2}\left((p\cdot p')^2+\frac{m^2m'^2}{\delta+2}\right),\quad
C=-p\cdot p'.
\label{ABC}
\ee
Upon differentiation with respect to $\tau$ and integration over $k^0$, using that in the chosen Lorentz frame
$\delta(k\cdot p')=\delta(m'k^0)$, one obtains
 \be
 \delta \zt^\mu(\tau)= \frac{2}{ \pi^2 m'm M_{*}^{2+\delta}V}\int\sum_n \left.
 \frac{ \e^{ik\cdot b-ik\cdot p\tau/m}} { {\bf k}^2+p_T^2 }Q^\mu\right|_{k^0=0}
 d^3k.
 \label{vel}
 \ee

To calculate the momentum transfer in (\ref{tdef}) we need the
 asymptotic values of $\delta\zt^\mu$ as $\tau\to\pm\infty$, which one computes next. In the
 chosen Lorentz frame the exponent in (\ref{vel}) is
 \be
 k\cdot b-k\cdot p\tau/m= k_x b-k_z v\gamma\tau.
 \ee
(a) Start with the integral over $k_z$ in (\ref{vel}). Define
 \be
 I^\mu(\tau)\equiv \int_{-\infty}^{\infty} \frac{Q^\mu\e^{-ik_z \gamma v
 \tau}}{k_z^2+\varkappa^2}dk_z,\quad \varkappa^2=k_x^2+k_y^2+p_T^2.
 \ee
The terms $B$ and $C$, as well as the one proportional to $Ak_z$ vanish in the limit $\tau\to\pm\infty$, because
 \be
 \int_{-\infty}^{\infty} \frac{ \e^{-ik_z \gamma v
 \tau}}{k_z^2+\varkappa^2}dk_z=\frac{\pi}{\varkappa}\e^{-\varkappa \gamma v
 |\tau|}
 \to 0 \; ,\; {\rm as} \;\;\tau\to\pm \infty.
 \label{kazed}
 \ee
The term proportional to $Ak_0$ is zero, because it is evaluated at
$k_0=0$. The $y$ component vanishes by parity. So, the only
component left is the one proportional to $Ak_x$. In this case,
parity implies that only the sine part
 of the exponential contributes. Use
 \be
 \lim_{\tau\to\pm\infty} \frac{\sin(k_z \gamma v\tau)}{k_z \gamma
 v}=\pm\pi\delta(k_z \gamma v ),
 \ee
 to perform the $k_z$ integration. Then, insert $A$
from (\ref{ABC}) to obtain
 \be
 \lim_{\tau\to\pm\infty}I^x(\tau)=\pm\frac{\pi k_x}{2\gamma v m \varkappa^2}
 \left((p\cdot p')^2-\frac{m^2m'^2}{\delta+2}\right).
 \ee
(b) Next, denote by $K^2=k_y^2+p_T^2$ and perform the integral over
$k_x$
 \be
 \int_{-\infty}^{\infty} \frac{k_x \e^{ik_x b}}{k_x^2+K^2}d k_x =
\pi\e^{-Kb}.\label{kax}
 \ee
(c) Insert this into (\ref{vel}) and replace the sum over KK
modes by a continuous integration
 \be
 \sum_n\to \frac{V S_{\delta-1}}{(2\pi)^\delta } \int_0^\infty p_T^{\delta-1}dp_T,\quad
 S_{\delta-1}=\frac{2\pi^{\delta/2}}{\Gamma(\delta/2)},
 \ee
to obtain
 \be
 \delta \zt^x(\pm\infty)=\pm\frac{1}
 {2^{2+\delta}\pi^{\delta/2}m'm^2\gamma v \Gamma(\delta/2)M_{*}^{2+\delta}}
 \left((p\cdot p')^2-\frac{m^2m'^2}{\delta+2}\right)
 \int_{-\infty}^\infty dk_y\int_0^\infty \e^{-Kb}\; p_T^{\delta-1}d p_T .
 \ee
(d) To perform the remaining integrations pass to polar coordinates
$p_T=K\cos\alpha,\;\;k_y=K\sin\alpha,\quad p_T^{\delta-1}dk_y
dp_T=K^\delta
  d K \cos^{\delta-1}\alpha d\alpha$,
 and integrate over $K$ from zero to infinity and over $\alpha$ from
 $-\pi/2$ to $\pi/2$. The asymptotic values of the velocity of $m$ in the field of $m'$ are then:
 \be
\delta \zt^x(\pm\infty)=\pm\frac{2 \Gamma(\delta/2+1)}{\pi^{\delta/2
}b^{\delta+1}m'm^2\gamma v M_{*}^{2+\delta }} \left((p\cdot
p')^2-\frac{m^2m'^2}{\delta+2}\right).
 \ee
 Inserting this into (\ref{tdef}) we find for the square of the
 momentum transfer
 \be
 -t= 2^{4}\Gamma^2(\delta/2+1)\frac{m^2m'^2}{\pi^{\delta }\gamma^2 v^2M_*^2}
 \frac1{(M_* b)^{2(\delta+1)}}\left(\frac{(p\cdot p')^2}{m^2m'^2}-\frac{1}{\delta+2}\right)^2.
 \ee
In the ultrarelativistic limit $\gamma\gg 1, v \simeq 1,$ one has
$s=2p\cdot p'\gg mm'$ and the above expression simplifies to:
 \be
 -t= \frac{2^{ 2}\Gamma^2(\delta/2+1)}{\pi^{\delta }(M_* b)^{2(\delta+1)}} \, \frac{s^2}{M_*^2}.
 \ee
Finally, the differential cross-section, defined as usual by
$d\sigma=2\pi b db$, is
  \be
 \frac{d\sigma}{dt}=\frac{1}{(\delta+1)(-M^2_*t)}
 \left(-4\pi \Gamma^2(1+\delta/2)\frac{s^2}{M^2_*t}\right)^{1/(\delta+1)}.
  \ee
In particular, for $\delta=0$
 \be
 \frac{d\sigma}{dt}=\frac{4\pi G_4^2 s^2}{t^2},
   \ee
which coincides with the well-known formula for small angle scattering of General Relativity \cite{DeSc}.

The scattering angle $\theta$ is given by
$\tan\theta=\sqrt{-t}/m\gamma v$, thus, small scattering angles mean
$|t|\ll m^{2}\gamma^{2}v^{2}$, which for ultrarelativistic
velocities gives the range of validity of our approximation \be
(M_{\ast}b)^{\delta+1}\gg\frac{m'}{M_{\ast}}. \ee

\section{Relation to the eikonal approximation}
Consider the elastic scattering of massive scalar particles on the
brane in the high-energy limit $s\gg m^2$. The Born amplitude
contains the $t$-channel propagator involving the sum over
Kaluza-Klein modes. Passing to the continuous integration over the
momentum of gravitons ${\bf p}_T$ in extra dimensions we have
\cite{GRW}:
 \be
 {\cal M}_{\rm Born}(s,t)=\frac{s^2\varkappa_D^2}{2 (2\pi)^\delta}\int
 \frac{d^{\delta}p_T}{-t+p_T^2}.
 \ee
 This integral in the general case requires a UV cutoff. An
 alternative way to get the final amplitude for the small-angle high-energy scattering
 is to use the eikonalized form of the amplitude \cite{Em,GRW,GuSj}
\be
{\cal M}_{\rm eik}(s,t)=2is\int\e^{i{\bf q\cdot b}}\left(1-\e^{i\chi(s,b)}\right)d^2b,
\label{eik}
 \ee
where the two-dimensional vectors ${\bf q,\,b}$ lie in the
transverse plane, with $\bf b$ the impact parameter vector. The
transverse component $\bf q$ of the momentum transfer in this
approximation satisfies ${\bf q}^2 \approx -q^\mu q_\mu$, so that
$t\simeq -{\bf q}^2$. This expression in the usual four-dimensional
theory corresponds to summation of the ladder and crossed-ladder
diagrams (for a detailed calculation within the quantized linearized
General Relativity see \cite{KaOr}). Actually, this involves UV
divergent loop diagrams, but it can be shown that the leading
contribution in the high-energy limit is independent on the cutoff.
In the ADD linearized gravity the situation is believed to be the
same, though no explicit analysis is available. Therefore our
classical calculation provides an independent check of the
applicability of the eikonal approximation in the ADD framework.

The Born amplitude corresponds to the first term in the expansion of
the exponential in (\ref{eik})
 \be
 {\cal M}_{\rm Born}(s,t)=2s\int\e^{i{\bf q\cdot b}} \chi(s,b) d^2
 b,\label{Born}
 \ee
and is used to extract the eikonal $\chi$ as its inverse Fourier-transform
 \be
 \chi(s,b) =\frac1{2s}\int\e^{-i{\bf q\cdot b}} {\cal M}_{\rm Born}(s,t)\; \frac{d^2
 q}{(2\pi)^2}
 \label{eikinvers}.
 \ee
 Notice, that although the Born amplitude itself may be divergent, the integral
(\ref{eikinvers}) is finite {\it if one first integrates over $q$,
but not $p_{T}$}. Indeed, choose the coordinates as in the previous
section to write
 \be
 \chi(s,b) =\frac{s\varkappa_D^2}{4(2\pi)^{\delta+2}}\int \e^{-i{  q_xb}}
 \frac{dq_x dq_y dp_T^\delta}{q_x^2+\varkappa^2},\quad
 \varkappa^2=q_y^2+p_T ^2,
 \ee
and integrate first over $q_{x}$ using a contour integration  which
gives an exponential factor cutting  the potentially divergent
integral over $p_{T}$. Then, integrate over $q_y=\varkappa
\cos\alpha$ and $p_T=\varkappa\sin\alpha$
 to obtain
 \be
 \chi(s,b) =\frac{s\varkappa_D^2}{4(2\pi)^{\delta+2}}\int_0^\infty d\varkappa
 \int_{-\pi/2}^{\pi/2} d\alpha \,
 \e^{-\varkappa b} S_{\delta-1}\cos^{\delta-1}\alpha \,\varkappa^\delta
 =\left(\frac{b_c}{b}\right)^\delta,
 \label{eikexpli}
  \ee
where
 \be
 \label{bece}
 b_c\equiv \frac1{\sqrt{\pi}}\left( \frac{\varkappa_D^2\Gamma(\delta/2)\,s}
 {16\pi}\right)^{1/\delta}.
 \ee
Then, the eikonal amplitude (\ref{eik}) becomes
\be
{\cal M}_{\rm eik}(s,t)=4\pi is \int J_0(qb)\left(1-\e^{i
\chi(s,b)}\right)b db.
 \ee
The unity in the parenthesis gives no contribution. In the remaining
part and in the regime $q b_c\gg 1$ of interest here, one may
replace the Bessel function by its asymptotic to obtain
 \be
 {\cal M}_{\rm
 eik}(s,t)=2\pi i s \int_0^\infty \sqrt\frac{2b}{\pi q}
 \left[\e^{i\psi_{+}(q,b)-i\pi/4}+ \e^{i\psi_{-}(q,b)+i\pi/4}\right]db,
 \label{eikoji}
 \ee
 where
  \be
\psi_{\pm}(q,b)=\pm qb+\left(\frac{b_c}{b}\right)^\delta.
 \label{psi}
 \ee
This can be evaluated using the stationary phase method.  The exponent of the second term
having no critical points of first order in the
domain of integration, is ignored. The stationary point
for the first exponent  is
 \be
 \frac{d\psi_{+}}{db}\Big|_{b_s}=0,\quad b_s =\left(\frac{\delta
 b_c^\delta}{q}\right)^{1/(\delta+1)},\quad
 \psi''_{+}(b_s)=\frac{\delta (\delta+1)b_c^{\delta}}{b_s^{\delta+2}} \, ,
 \ee
and upon integration
\begin{align}
 {\cal M}_{\rm
 eik}(s,t)=\frac{4 \pi i s }{\sqrt{\delta(\delta+1)}} \frac{b_s^{(\delta+3)/2}}{b_c^{\delta/2}} \e^{i q
 b_s}=\frac{4 \pi i s \delta^{1/(\delta+1)}}{\sqrt{\delta+1}}
 \frac{b_c^{\frac{\delta}{\delta+1}}}{q^{\frac{\delta+2}{\delta+1}}} \e^{i
 qb_s}.
\end{align}
Substituting $b_c$ from (\ref{bece}) one obtains
\begin{align}
 {\cal M}_{\rm
 eik}(s,t)=\frac{4\sqrt{\pi} s\e^{i(qb_s-\pi/2)}}{q\sqrt{\delta+1}} \!\left(\!\frac{\varkappa_D^2 s
 \Gamma(\delta/2+1)}
 {8 \sqrt{\pi} q }\!\right)^{\!\!\frac{1}{\delta+1}}\!=\frac{4\sqrt{\pi} s\e^{i(qb_s-\pi/2)}}{q\sqrt{\delta+1}}
 \!\left(\!\frac{2 \sqrt{\pi} s \Gamma(\delta/2+1)}
 {M_*^{\delta+2} q }\!\right)^{\!\!\frac{1}{\delta+1}}.
 \label{stat}
\end{align}
Note, that although a few intermediate steps are singular for $\delta=0$ and seem to require a separate discussion,
the final formula is valid for $\delta=0$ as well.

The corresponding cross-section reads
  \be
 \frac{d\sigma_{\rm eik}}{dt}=\frac{1}{16\pi s^2}|{\cal M}_{\rm eik}|^2=\frac{1}{(\delta+1) M^2_*|t|}
 \left(4\pi \Gamma^2(1+\delta/2)\frac{s^2}{M^2_*|t|}\right)^{1/(\delta+1)}.
  \ee
 As advertised, it is identical to our classical result.

\section{Conclusions}
A purely classical calculation was presented of the high energy
elastic scattering cross section in the ADD scenario. Our approach
is entirely free of the ambiguities associated with the divergence
of the Born amplitude with the virtual graviton exchange typical for
ADD. Ultrarelativistic small angle gravitational collision in four
dimensions is a special case, in which it agrees with 't Hooft' s
result, which in turn coincides with the Born quantum cross-section.
In the presence of extra dimensions it was shown that the lowest
order small angle classical approximation reproduces the essentially
non-perturbative result of the quantum eikonal calculation in the
saddle-point approximation. Thus, the classical computation in the
above kinematical regime is non-trivial, unambiguous as well as
reliable and, therefore, worth applying to other processes like
bremsstrahlung \cite{br}.

\section*{Acknowledgments}
Work supported in part by the EU grants INTERREG IIIA (Greece-Cyprus),
MRTN-CT-2004-512914 and the FP7-REGPOT-2008-1-CreteHEPCosmo-228644.
DG and PS are grateful to the Department of Physics of the
University of Crete for its hospitality in the early stages of this work.
Their work was also supported by the RFBR under the project
08-02-01398-a.


\end{document}